\documentstyle[12pt]{article}

\newcommand{\be}{\begin{equation}}
\newcommand{\ee}{\end{equation}}
\newcommand{\ba}{\begin{array}{c}}
\newcommand{\ea}{\end{array}}
\newcommand{\bqa}{\begin{eqnarray}}
\newcommand{\eqa}{\end{eqnarray}}

\begin{document}

\begin{titlepage}
\begin{flushright}
PSI-PR-95-5\\
\end{flushright}
\vspace{3cm}
\begin{center}

\begin{Large}
{\bf A comment on  the anomalous $Zt\bar t$ couplings and the
$Z\rightarrow b\bar b$ decay}
\end{Large}
\vspace*{1.5cm}

{\bf Denis Comelli}\footnote{
supported by the "Ministerio de Education y Ciencia de Espana" }\\
\vspace*{0.3cm}

University of Valencia 46100 Burjassot (Valencia),
Spain\footnote{e--mail: comelli@evalvx.ific.uv.es}\\

\vspace*{0.5cm}
and\\
\vspace*{0.5cm}
{\bf Hanqing Zheng}\\
\vspace*{0.3cm}
Theory group, P. Scherrer Institute, 5232 Villigen PSI,
Switzerland\footnote{e--mail: zheng@cvax.psi.ch}\\
\end{center}
\vspace*{1.5cm}

\begin{abstract}
By reanalyzing the influence of the anomalous $Z\bar t t$ couplings
on the $Z\rightarrow b\bar b$ decay process,
we pointed out the ambiguity in the conventional treatment of
the effective Lagrangian approach, because of the possible existence
of large contributions given by constant terms
beyond the leading cutoff dependent term.
\end{abstract}
\vfill
\vfill
\phantom{p}
\vfill
\end{titlepage}

The remarkable experimental result from LEP
at CERN on the $Z\rightarrow
b\bar b$ decay indicates that \cite{blondel},
\be\label{exp}
R_b={\Gamma_b\over \Gamma_{hadrons}} = 0.2192\pm 0.0018  \ ,
\ee
which is about 3$\sigma$
away from the theoretical value predicted by
the standard model \cite{bardin}, for the supported value
of the top quark mass, $m_t=174\pm 20$GeV \cite{cdf}. Intuitively,
one may think that such a heavy top quark
mass  at the order of the Fermi scale may be
useful in probing the electroweak
symmetry breaking sector. As discussed in ref.~\cite{zhang}, if the
electroweak symmetry is broken dynamically one
would expect, in general,
the violation of the fermion--gauge boson coupling universality.
The anomalous fermion--gauge boson coupling may naturally be
expected at the
order of  $\sim \sqrt{m_im_j}/v$ \cite{pe,frigeni}, where $i,j$ refer to
the two fermions involved. Therefore the
most important deviation from the
standard model may be
reflected in the anomalous $Zt\bar t$ couplings, which are
characterized by two parameters, $\kappa_L$
and $\kappa_R$, corresponding
to the anomalous $Z$ coupling to the left--hand current and the
right--hand current, respectively. The
observed discrepancy between the experimental
value of $R_b$ and the standard model prediction may therefore be
 explained
by the nonvanishing of these anomalous couplings
\cite{pe,frigeni,yuan}.

The effective $Zt\bar t$ Lagrangian obtained from a
gauge invariant non-linear chiral Lagrangian
for the Goldstone boson--top
quark interactions, in the unitary gauge, can simply be written as,
\be\label{L}
L={g\over 2c}\bar t\gamma_\mu\{P_L(1+\kappa_L) + P_R\kappa_R
-{8\over 3}s^2\} t Z^\mu \ .
\ee
In
order to read off the physical meaning of the non-renormalizable
Lagrangian, it is appropriate to replace the divergence by a physical
cutoff which is regarded as the scale of new physics,
in a gauge invariant way. Following previous calculations we use
dimensional regularization scheme that is to interpret
$1/\epsilon$ by $\ln(\Lambda)$\cite{cut}.

The physical observables at Z threshold
can be quoted in four quantities,
$\epsilon_{1,2,3}$ and $\delta_{bV}$ (for the definitions of these
parameters, see ref.~\cite{alt}, \cite{verz}).
The expressions of these quantities as  functions of the anomalous
$Zt\bar t$ couplings and the cutoff $\Lambda$ were
previously obtained,
in the leading term approximation of the $m_t^2$ expansions, with the
constant terms beside the $\ln\Lambda$ term systematically neglected.
Using the new LEP data (see ref.~\cite{altar}, for the measure of
$\delta_{bV}$  see ref.~\cite{renard})
\be\label{eps1}
-2.4\times 10^{-3} < \epsilon_1 < 2\times 10^{-3} \ ,
\ee
\be
-5.8\times 10^{-3}< \epsilon_2 < 4.8\times 10^{-3}\ ,
\ee
\be
-5.2\times 10^{-3}< \epsilon_3 < 0.8\times 10^{-3}\ ,
\ee
\be\label{4}
30.4\times 10^{-3}< \delta_{bV} < 52.4\times 10^{-3}\ .
\ee
we redraw the 1$\sigma$  plot as shown in fig.~1.
As can be seen from fig.~1, the
constraints on $\kappa_L$ and $\kappa_R$
from the experimental values of the four observables
are in severe disagreement with each other. In the effective
Lagrangian approach to exploring the electroweak physics,
one should in general write down the whole set of the dimension
six operators,
including the anomalous couplings among the  gauge bosons
(induced from the Goldstone boson--gauge boson
interactions in the unitary
gauge) \cite{hagiwara} and the anomalous
$Wt\bar b$ couplings \cite{yuan}.
Both the two types of anomalous couplings contribute to the epsilon
quantities.
Assuming a CP conserving effective interaction, in the
former case there are 9 parameters
while in the later case there are two,
$\kappa^{CC}_L$ and $\kappa^{CC}_R$. Including more parameters in the
1$\sigma$ plot one expect that there can be a region in the parameter
space
fitting the experimental values. If one restricts to
the anomalous gauge boson couplings only, it is
shown \cite{renard}
that influence on $\delta_{bV}$ comes essentially from
only one operator in the effective Lagrangian.
This approach, orthogonal to that
of considering the anomalous $Z t\bar t$ couplings only, emphasizes
different possible physics beyond the
standard model as assuming the top
quark playing no special role except that it has a large mass.
Adding anomalous $Wt\bar b$ couplings, is helpful in solving the
difficulties shown by fig.~1.
However, the anomalous $Wt\bar b$ couplings
will contribute to the $b\rightarrow s\gamma$ process by violating the
GIM  mechanism and may therefore receive strong constraints.
A most general analysis, combining all the possible effects on the
$Z\rightarrow b\bar b$ process, will however  be less predictive.

One may conclude that the "minimal" fit in fig.~1
to the experimental results is inadequate, and
therefore more  parameters have to be added.
However  this
simple approach is still possibly being self-consistent.
The way is that in the previous calculation all
the constant contributions (with respect to the term proportional to
$\ln(\Lambda/m_t))$ induced by
the anomalous couplings are neglected.
In principle constant terms should also contribute to
physical observables and in practice their contribution
can be large numerically, as already known in
chiral perturbation theory for hadron interactions.
There is no a priori reason to neglect these
effects. However the problem
is that no unambiguous way exists in estimating them.
In a cutoff dependent
theory sub-leading terms are usually regularization
procedure dependent. This can partly be considered as
 a reflection of the ambiguity
in  determining the explicit
value of the cutoff parameter. In a renormalizable
theory, all  of these
uncertainties are absorbed into the
renormalized coupling constants which are
identified as the experimentally observed values. Finite quantities,
such as
the anomalous magnetic moment of the electron, are cutoff independent
and are
therefore free of the ambiguity caused by the explicit dependence
of the cutoff
parameter\footnote{Terms which are vanishing when sending
the cutoff to infinity are
also regularization procedure dependent,
even in renormalizable theories. This means if
we consider the renormalizable theory as an
"effective" theory, i.e., keeping
the cutoff parameter finite, there are also
ambiguities, at the $O(1/\ln\Lambda)$ or $O(1/\Lambda)$ level.}.
Once the effective theory is embedded
into the underlying renormalizable theory
the cutoff parameter is replaced by some heavy mass scale which are
physical
observables \cite{cut}.
The constant terms in the results given by the
effective Lagrangian approach will be shifted because there are also
contributions of such
kind from high energy sector which are practically unknown
(As argued in ref.~\cite{cut} only the coefficient of
the $\ln\Lambda$ term can be fixed).
{}From these terms obtained from the effective Lagrangian
we may at best
know something about  the order of  magnitude
of these finite corrections.

In the present case, we list the full expressions of the
the above epsilons,
including the constant terms, in the leading order of $m_t^2$ expansions:
$\epsilon_i=\epsilon_i^{SM}+\epsilon_i'$
and $\delta_{bV}= \delta_{bV}^{SM}+
\delta_{bV}'$ ($\epsilon_i^{SM}$
and $\delta_{bV}^{SM}$ are the standard model
contributions),

\be \label{log}
\epsilon_1' =
{3G_Fm_t^2\over 4\sqrt{2}\pi^2}
[2(k_R-k_L)-(k_R-k_L)^2]\ln({\Lambda^2\over m_t^2})\ ,
\ee
\be
\epsilon_2' =
-{G_Fm_W^2\over \sqrt{2}\pi^2}[({1\over 2}k_L+{k_L^2+k_R^2\over 4})
\ln({\Lambda^2\over m_t^2})
+ {k_R-k_L\over 2}-{(k_R-k_L)^2\over 4}]\ ,
\ee
\be
\epsilon_3' =
{G_Fm_t^2\over 2\sqrt{2}\pi^2}
[ ({2\over 3}k_R-{1\over 3}k_L-{k_L^2+k_R^2\over 2})
\ln({\Lambda^2\over m_t^2})
+k_L-k_R+{(k_L-k_R)^2\over 2}]\ ,
\ee
\be
\delta_{bV}' =
{G_Fm_t^2\over\sqrt{2}\pi^2}[(k_L-{1\over 4}k_R)
\ln({\Lambda^2\over m_t^2})+{1\over 2}k_L + {1\over 8}k_R]\ .
\ee
There are still ambiguities in the above expressions since
a redefinition of the cutoff parameter will change the constant terms.
We may remove the part in  each constant term
which is proportional to the
coefficient of the corresponding
$\ln(\Lambda/m_t)$ term. In other words, this
part is being absorbed by the cutoff dependent term and the remaining
quantity, by definition, is not influenced by the change of the cutoff
parameter. Then we may rewrite the above expressions as:
\be \label{finite}
\epsilon_1' =
{3G_Fm_t^2\over 4\sqrt{2}\pi^2}
[2(k_R-k_L)-(k_R-k_L)^2]\ln({\Lambda^2\over m_t^2})
\ ,\ee
\be
\epsilon_2' =
-{G_Fm_W^2\over \sqrt{2}\pi^2}[({1\over 2}k_L+{k_L^2+k_R^2\over 4})
\ln({\Lambda^2\over m_t^2}) + {k_R\over 2}+{k_Rk_L\over 2}]\ ,\ee
$$
\epsilon_3' =
{G_Fm_t^2\over 2\sqrt{2}\pi^2}
[ ({2\over 3}k_R-{1\over 3}k_L-{k_L^2+k_R^2\over 2})
\ln({\Lambda^2\over m_t^2}) $$
\be
+{2\over 5}k_L+{1\over 5}k_R-{2\over 5}(k_L^2+k_R^2)-k_Lk_R]\ ,\ee
\be
\delta_{bV}' =
{G_Fm_t^2\over\sqrt{2}\pi^2}[(k_L-{1\over 4}k_R)
\ln({\Lambda^2\over m_t^2})
+{4\over 17}({1\over 4}k_L + k_R)]\ . \label{eps2}
\ee

The effects of these additional terms can be
seen by comparing fig.~1 with
 fig.~2. We find that their effects are large. In the present example,
they  improve impressively the discrepancies
shown in fig.~1. Therefore it may
still be possible to explain the 3$\sigma$ discrepancies between the
experimental values and the standard model prediction of $R_b$ in terms
of only two anomalous parameters.
Certainly, we are not able  to demonstrate
this two parameter fit is valid when taking the constant contributions
into account.
What one can only conclude from the above example is that
the conventional treatment to the effective Lagrangian approach to
exploring the possible physics beyond the standard model suffers from
ambiguities: the uncontrollable constant
terms can be large in magnitude in some cases, and
in the worst situation  one may imagine (although not natural),
can destroy  results
obtained by simply ignoring them.

The unambiguous way in exploring the possible anomalous effects,
can be like that have been done in chiral perturbation
theory\cite{leutwyler}:
including all
possible terms at dimension six,
cancelling all the divergence by counter
terms, and bet that there are still
predictions left. However since too many
parameters are involved in the present case, and in general,
we suspect that the useful informations
can be obtained are rather limited.

We thank Fred Jegerlehner and Claudio Verzegnassi for helpful discussions.

\begin{newpage}
Fig.1: Contourplot of $\epsilon_1',\epsilon_2',\epsilon_3', \delta_{bV}'$
including only the $\log \Lambda$ terms
 in function of $ k_L, k_R$  with the
constraints of eqs.~(\ref{eps1}) to (\ref{4})
and with $m_t=174$ GeV, $\Lambda=1$ TeV, $m_H=300$ GeV.
The interval of variation of the $\epsilon_i', \delta_{bV}'$ parameters
have been found imposing the constraints at 1 $\sigma$:
$$
\epsilon_i'^{max}=(\bar{\epsilon_i}-\epsilon_i'^{SM})+ \epsilon_i^{max}
$$
$$
\epsilon_i'^{min}=(\bar{\epsilon_i}-\epsilon_i'^{SM})+ \epsilon_i^{min}
$$
where $\bar{\epsilon_i}$ is the
experimental mean value, $\epsilon_i^{SM}$
is the corresponding theoretical value, whereas $\epsilon_i^{max(min)}$
are the extremal values at 1 $\sigma$ away  from
$\bar{\epsilon_i}$.

Fig.2:  The same as before but with the
$\epsilon_i', \delta_{bV}'$  including
also the finite  contributions.
\end{newpage}

\begin{thebibliography}{9}
\bibitem{blondel}
A.~Blondel, CERN-PPE/94-133 (1994).
\bibitem{bardin}
A.~Akhundov, D.~Bardin and T.~Rieman,
Nucl. Phys. {\bf B276} (1986) 1;\\
W.~Beenakker and W.~Hollik, Z.~Phys. {\bf C40} (1988) 11;\\
B.~W.~Lynn and R.~G.~Stuart, Phys. Lett. {\bf B252} (1990) 676;
J.~Bernabeu, A.~Pich, and A.~Santamaria, Nucl.
Phys. {\bf B363} (1991) 326.
\bibitem{cdf}
The SLD Collaboration, K.~Abe et al.,
Phys. Rev. Lett. {\bf 73} (1994) 225.
\bibitem{zhang}
R.~D.~Peccei and X.~M.~Zhang, Nucl. Phys. {\bf B337} (1990) 269.
\bibitem{pe}
R.~D.~Peccei S.~Peris and X.~M.~Zhang,
Nucl. Phys. {\bf B349} (1991) 305.
\bibitem{frigeni}
M.[B~Frigeni and R.~Rattazzi, Phys. Lett. {\bf B269} (1991) 412.
\bibitem{yuan}
E.~Malkawi and C.~P.~Yuan, Phys. Rev. {\bf D50} (1994) 4462.
\bibitem{cut}
C.~P.~Burgess and D.~London, Phys. Rev. {\bf D48} (1993) 4337.
\bibitem{alt}For the definition of
these $\epsilon$ quantities, see
G.~Altarelli, R.Barbieri,  Nucl. Phys. {\bf B369} (1992) 3.
\bibitem{verz}A.Blondel, A.Djouadi, C.Verzegnassi, Phys. Lett.
{\bf B293} (1992) 253.
\bibitem{hagiwara}K.~Hagiwara, S.~Ishihara,
R.~Szalapski and D.~Zeppenfeld,
Phys. Rev. {\bf D48} (1993) 2182.
\bibitem{renard}F.~M.~Renard and C.~Verzegnassi,
Preprint CERN-TH.7376/94.
\bibitem{leutwyler}J.~Gasser and H.~Leutwyler,
Nucl. Phys. {\bf B250}
(1985) 465.
\bibitem{altar}G.~Altarelli, preprint CERN-TH 7319/94.
\end{thebibliography}
\end{document}